\documentclass[conference]{IEEEtran}

\usepackage[utf8]{inputenc}
\usepackage[T1]{fontenc}
\usepackage{url}
\usepackage{cite}

\usepackage{booktabs}
\usepackage{multirow}
\usepackage{threeparttable}
\usepackage{array}

\usepackage{amsmath}
\usepackage{amssymb}
\usepackage{amsfonts}
\usepackage{amsthm}
\usepackage{mathrsfs}
\usepackage{nicefrac}

\usepackage{graphicx}

\usepackage[caption=false,font=footnotesize]{subfig}

\usepackage{microtype}
\usepackage{siunitx}
\usepackage{xcolor}
\usepackage{enumerate}
\usepackage{cleveref}

\usepackage{fancybox}
\usepackage{calc}

\definecolor{customgreen}{RGB}{204,238,208}
\theoremstyle{plain}

\theoremstyle{definition}

\theoremstyle{remark}

\crefname{Theorem}{Theorem}{Theorems}
\Crefname{Theorem}{Theorem}{Theorems}
\crefname{Proposition}{Proposition}{Propositions}
\Crefname{Proposition}{Proposition}{Propositions}
\crefname{Lemma}{Lemma}{Lemmas}
\Crefname{Lemma}{Lemma}{Lemmas}
\crefname{Corollary}{Corollary}{Corollaries}
\Crefname{Corollary}{Corollary}{Corollaries}
\crefname{Definition}{Definition}{Definitions}
\Crefname{Definition}{Definition}{Definitions}
\crefname{Remark}{Remark}{Remarks}
\Crefname{Remark}{Remarks}{Remarks}
\crefname{Assumption}{Assumption}{Assumptions}
\Crefname{Assumption}{Assumption}{Assumptions}

\begin{document}

\title{Carbon-Aware Compute--Power Scheduling for AI Data Centers with Microgrid Prosumer Operations}

\author{
\IEEEauthorblockN{
Johnny R. Zhang\IEEEauthorrefmark{1},
Gaoyuan Du\IEEEauthorrefmark{2},
Qianyi Sun\IEEEauthorrefmark{3},
Shiqi Wang\IEEEauthorrefmark{4},
Jiaxuan Li\IEEEauthorrefmark{2},
and Xian Sun\IEEEauthorrefmark{5}
}
\IEEEauthorblockA{
\IEEEauthorrefmark{1}Independent Researcher, Email: johnny.r.zhang97@gmail.com}
\IEEEauthorblockA{
\IEEEauthorrefmark{2}Amazon, Emails: gdu@amazon.com, lidjlz.170400@gmail.com}
\IEEEauthorblockA{
\IEEEauthorrefmark{3}Microsoft, Email: erisun@microsoft.com}
\IEEEauthorblockA{
\IEEEauthorrefmark{4}Meta, Email: swang151@fordham.edu}
\IEEEauthorblockA{
\IEEEauthorrefmark{5}Duke University, Email: xiansun@alumni.duke.edu}
}

\maketitle

\begin{abstract}
AI data centers are increasingly becoming tightly coupled compute--energy systems, where workload placement, cooling demand, electricity procurement, storage operation, and carbon emissions interact over time. This paper studies carbon-aware compute--power scheduling for geographically distributed AI data centers with microgrid prosumer capabilities. We propose a mixed-integer linear programming (MILP) framework that jointly schedules rigid training jobs, routes elastic inference workloads, dispatches local generation and battery storage, and manages bidirectional grid interaction under latency, continuity, power-balance, and carbon-budget constraints. The model captures two key features of emerging AI infrastructure: heterogeneous workload flexibility and site-level energy prosumer operation. Experiments on synthetic yet practically motivated instances show that the proposed joint MILP substantially improves total operational benefit over compute-only and energy-only baselines while reducing emissions. The results further indicate that inference-routing flexibility is a major source of value, battery storage provides useful temporal flexibility, and local-generation-rich settings are particularly favorable. The framework provides a tractable optimization abstraction for sustainable and grid-interactive AI data centers.
\end{abstract}

\begin{IEEEkeywords}
AI data centers, MILP, carbon-aware scheduling, microgrids
\end{IEEEkeywords}

%
\IEEEpeerreviewmaketitle

\section{Introduction}
\label{sec:introduction}

The rapid expansion of artificial intelligence (AI) services has substantially increased the scale and energy demand of modern data centers. Large-scale model training creates rigid, high-power workloads, while online inference creates latency-sensitive but geographically routable demand. At the same time, hyperscale AI operators are increasingly securing dedicated energy resources, deploying storage, and actively managing electricity procurement through power purchase agreements, co-located clean generation, and on-site energy assets. These trends suggest that AI data centers should no longer be modeled only as passive electricity consumers, but rather as integrated compute--energy systems with prosumer capabilities.

A central challenge is the tight coupling between computation and power system operation. Workload placement determines IT power consumption, which further induces cooling demand and affects grid imports, local generation usage, battery dispatch, and carbon emissions. Conversely, electricity prices, carbon intensity, renewable availability, and battery state-of-charge influence where and when AI workloads should be served. This bidirectional compute--power coupling is particularly important for AI infrastructure because training jobs are often non-preemptive and spatially indivisible, whereas inference workloads are more elastic but subject to latency and service-level constraints.

This paper develops a unified optimization framework for carbon-aware compute--power scheduling in geographically distributed AI data centers with microgrid prosumer capabilities. The proposed model jointly captures rigid training-job commitment, elastic inference routing, dynamic cooling overhead, local generation, battery storage, bidirectional grid interaction, and a system-wide carbon budget. The resulting problem is formulated as a mixed-integer linear program (MILP), enabling workload scheduling and energy dispatch to be optimized within a single tractable decision model.

The main contributions are as follows. First, we formulate AI data centers as microgrid prosumers whose compute and energy decisions are operationally coupled. Second, we develop a MILP model that integrates heterogeneous AI workloads, cooling-adjusted power demand, grid import/export decisions, battery dynamics, latency constraints, and carbon accounting. Third, we evaluate the proposed framework on synthetic yet practically motivated scenarios and compare it with decoupled compute-only, energy-only, no-battery, no-routing, and no-carbon baselines. The results quantify the value of joint scheduling and identify inference-routing flexibility, storage, and local generation as key drivers of operational performance.\footnote{This work was conducted independently and is unrelated to the authors' positions at their affiliated organizations. The views expressed are those of the authors and do not represent their affiliated organizations.}

\section{Related Work}
\label{sec:related_work}

This work connects three streams of literature. First, energy management in data centers has studied electricity-cost reduction, geographical load balancing, renewable-powered operation, and hybrid power-supply architectures~\cite{qureshi2009cutting,liu2011greening,goiri2013parasol,khosravi2024review}. Second, microgrid and prosumer optimization has developed model-predictive and mixed-integer formulations for coordinating local generation, storage, grid interaction, and energy trading~\cite{parisio2014mpc,misljenovic2023review,garcia2025milp}. Third, AI workload scheduling has addressed GPU-cluster management, distributed training placement, and training/inference scheduling in large-scale data centers~\cite{xiao2018gandiva,gu2019tiresias,ye2024survey,choudhury2024mast}.

This paper is also related to broader studies on cyber-physical production and product-service systems, where sensing, cloud platforms, quality control, and operational decision models are integrated to improve system-level performance. Prior work has studied IoT-enabled remote monitoring and control for manufacturing systems~\cite{guo2020online}, plant-wide sustainable quality-control mechanisms~\cite{guo2020sustainable}, task pricing in product-service platforms~\cite{guo2019task}, and product/service improvement prioritization based on importance-performance analysis~\cite{wu2018approach}. While these studies focus mainly on manufacturing and product-service operations, they share a common modeling perspective with this paper: operational decisions should be optimized jointly with the supporting cyber-physical infrastructure.

Carbon-aware computing is also closely related. Prior work has shown that spatial and temporal variation in grid carbon intensity can guide cloud load balancing and flexible datacenter workloads~\cite{zhou2013carbon,radovanovic2023carbonaware}, while recent systems study carbon-aware datacenter design and ML inference~\cite{acun2023carbonexplorer,li2023clover}. Compared with these studies, our focus is a unified optimization layer for AI prosumer data centers, where rigid training, elastic inference routing, cooling, local generation, storage, grid transactions, and carbon constraints are optimized jointly.
\section{System Model and Formulation}
\label{sec:model_formulation}

We consider a set of geographically distributed AI data-center sites
$i\in\mathcal I$ over time periods $t\in\mathcal T$. Each site has
computing resources, local generation, grid-interconnection capacity, and
battery energy storage. The workload consists of rigid training jobs
$j\in\mathcal J_{\rm tr}$ and elastic inference classes
$k\in\mathcal J_{\rm inf}$. Training jobs are modeled as non-preemptive and
spatially indivisible workloads, while inference demand can be routed across
sites subject to latency constraints. A central scheduler jointly determines
training commitment, inference routing, battery dispatch, grid import/export,
and carbon-aware energy usage.

Let $u_{i,j,t}\in\{0,1\}$ indicate whether training job $j$ runs at site $i$
in period $t$, and let $v_{i,j,t},w_{i,j,t}\in\{0,1\}$ denote startup and
shutdown decisions. Let $x_{i,k,t}\ge0$ be the amount of inference workload
class $k$ assigned to site $i$. Energy variables include grid import/export
$P^{\rm buy}_{i,t},P^{\rm sell}_{i,t}$, battery charging/discharging
$P^{\rm chg}_{i,t},P^{\rm dis}_{i,t}$, state of charge ${\rm SOC}_{i,t}$,
IT power $P^{\rm IT}_{i,t}$, and total facility power $P^{\rm tot}_{i,t}$.
The binary variables $y^{\rm buy}_{i,t},y^{\rm sell}_{i,t}$ indicate grid
import/export states.

The objective maximizes total net operational benefit:
\begin{align}
\max \sum_{t\in\mathcal T}\sum_{i\in\mathcal I}
\Bigg[
&\sum_{k\in\mathcal J_{\rm inf}}
\big((R_k-C_k^{\rm gpu})-\gamma^{\rm sla}\tau_{i,k}\big)x_{i,k,t}
\nonumber\\
&+\lambda_t^{\rm sell}P^{\rm sell}_{i,t}
-\lambda_t^{\rm buy}P^{\rm buy}_{i,t}
-C^{\rm deg}\big(P^{\rm chg}_{i,t}+P^{\rm dis}_{i,t}\big)
\Bigg].
\label{eq:obj}
\end{align}

Workload decisions determine IT and total facility power:
\begin{align}
P^{\rm IT}_{i,t}
&=\sum_{j\in\mathcal J_{\rm tr}}P^{\rm tr}_j u_{i,j,t}
+\sum_{k\in\mathcal J_{\rm inf}}P^{\rm inf}_k x_{i,k,t},
&&\forall i,t, \label{eq:it_power}\\
P^{\rm tot}_{i,t}
&=(1+\alpha_{i,t})P^{\rm IT}_{i,t},
&&\forall i,t. \label{eq:total_power}
\end{align}

Site-level energy balance and grid interaction are modeled by
\begin{align}
P^{\rm loc}_{i,t}+P^{\rm buy}_{i,t}+P^{\rm dis}_{i,t}
&=
P^{\rm tot}_{i,t}+P^{\rm sell}_{i,t}+P^{\rm chg}_{i,t},
&&\forall i,t, \label{eq:energy_balance}\\
P^{\rm buy}_{i,t}
&\le C_i^{\rm grid}y^{\rm buy}_{i,t},
&&\forall i,t, \label{eq:grid_buy}\\
P^{\rm sell}_{i,t}
&\le C_i^{\rm grid}y^{\rm sell}_{i,t},
&&\forall i,t, \label{eq:grid_sell}\\
y^{\rm buy}_{i,t}+y^{\rm sell}_{i,t}
&\le 1,
&&\forall i,t. \label{eq:grid_exclusive}
\end{align}

Rigid training jobs satisfy assignment and continuity constraints:
\begin{align}
\sum_{i\in\mathcal I}u_{i,j,t}
&\le 1,
&&\forall j,t, \label{eq:training_unique}\\
u_{i,j,t}-u_{i,j,t-1}
&=v_{i,j,t}-w_{i,j,t},
&&\forall i,j,t, \label{eq:state_transition}\\
\sum_{\tau=t}^{t+U_j^{\min}-1}u_{i,j,\tau}
&\ge U_j^{\min}v_{i,j,t},
&&\forall i,j,t. \label{eq:min_up}
\end{align}

Elastic inference demand must be fully served within latency-feasible routes:
\begin{align}
\sum_{i\in\mathcal I}x_{i,k,t}
&=D_{k,t},
&&\forall k,t, \label{eq:inference_demand}\\
x_{i,k,t}
&=0,
&&\forall t\ \text{if } \tau_{i,k}>\tau^{\max}. \label{eq:latency_cutoff}
\end{align}

Battery operation and carbon accounting are given by
\begin{align}
{\rm SOC}_{i,t}
&={\rm SOC}_{i,t-1}
+\eta^{\rm chg}P^{\rm chg}_{i,t}
-\frac{1}{\eta^{\rm dis}}P^{\rm dis}_{i,t},
&&\forall i,t, \label{eq:soc}\\
\underline{\rm SOC}_i
&\le {\rm SOC}_{i,t}\le \overline{\rm SOC}_i,
&&\forall i,t, \label{eq:soc_bound}\\
0\le P^{\rm chg}_{i,t}
&\le \overline P_i^{\rm chg},
\quad
0\le P^{\rm dis}_{i,t}\le \overline P_i^{\rm dis},
&&\forall i,t, \label{eq:battery_bounds}\\
\sum_{t\in\mathcal T}\sum_{i\in\mathcal I}
\rho^{\rm CO_2}_{i,t}P^{\rm buy}_{i,t}
&\le E^{\max}. \label{eq:carbon_budget}
\end{align}

Together with the binary domains of
$u,v,w,y^{\rm buy},y^{\rm sell}$ and the nonnegativity of continuous
power and workload variables, \eqref{eq:obj}--\eqref{eq:carbon_budget}
define a mixed-integer linear program. The formulation is linear because
cooling coefficients, electricity prices, carbon intensities, workload
parameters, and storage efficiencies are treated as exogenous inputs.

\section{Computational Approach}
\label{sec:computational_approach}

The resulting model is a structured MILP and can be solved directly by
branch-and-bound or branch-and-cut solvers. In the experiments, we use
Gurobi with a fixed time limit and relative MIP-gap tolerance. Before
optimization, latency-infeasible routing pairs are removed, and physical
bounds on grid and battery variables are tightened.

The problem is NP-hard. To see this, consider a single-site, single-period
special case with no inference demand, no battery storage, no grid selling,
no carbon constraint, and unit minimum up-time. The remaining problem is to
select a subset of training jobs with power requirements $P_j^{\rm tr}$
under a capacity budget while maximizing reward, which is the classical
0--1 knapsack problem. Hence, the full compute--power scheduling problem is
NP-hard.
Although the present paper solves deterministic finite-horizon MILP instances directly, larger online variants may require decomposition, stochastic approximation, or first-order iterative schemes. Recent work on general stochastic iteration frameworks provides a possible theoretical foundation for such extensions in learning and large-scale decision systems~\cite{zhang2025universal}.
\section{Experiments}
\label{sec:experiments}

We evaluate the proposed MILP on synthetic yet practically motivated AI
prosumer data-center instances. The default instance contains 3 sites, 24
hourly periods, 6 rigid training jobs, and 3 elastic inference classes. All
models are implemented in Python and solved by Gurobi with a 120-second time
limit and a 1\% relative MIP-gap tolerance. We compare the proposed joint
MILP with five baselines: compute-only scheduling, energy-only dispatch,
no-battery, no-routing, and no-carbon variants. The main metrics are total
objective value, emissions, runtime, and final optimality gap.

\subsection{Baseline Comparison}
\label{subsec:baseline_comparison}

Table~\ref{tab:overall_results} reports the default-instance results. The
proposed joint MILP achieves the highest objective value, $206{,}133.08$,
compared with $59{,}061.42$ for compute-only scheduling and $66{,}000.31$
for energy-only dispatch. It also reduces emissions from $1{,}978.47$ and
$1{,}918.35$ to $707.08$, respectively. These results show that optimizing
only the workload layer or only the energy layer leaves substantial value
unrealized.

The no-battery baseline remains competitive but has lower objective value
and higher emissions, indicating that storage provides useful but secondary
temporal flexibility. In contrast, the no-routing baseline performs worst,
with objective value $30{,}074.24$ and emissions $2{,}795.40$, confirming
that inference-routing flexibility is a primary source of value. The
no-carbon baseline is nearly identical to the proposed MILP, suggesting that
the default carbon budget is weakly binding.

\begin{table}[!t]
\centering
\caption{Default-instance comparison.}
\label{tab:overall_results}
\footnotesize
\begin{tabular}{lrrrr}
\toprule
Method & Obj. & Emis. & Time & Gap \\
\midrule
Joint MILP & 206133 & 707 & 0.060 & 0.000 \\
Compute-only & 59061 & 1978 & 0.066 & 0.000 \\
Energy-only & 66000 & 1918 & 0.020 & 0.000 \\
No-battery & 194166 & 843 & 0.027 & 0.008 \\
No-routing & 30074 & 2795 & 0.042 & 0.000 \\
No-carbon & 206133 & 706 & 0.040 & 0.000 \\
\bottomrule
\end{tabular}
\end{table}

\subsection{Sensitivity and Scalability}
\label{subsec:sensitivity_scalability}

\begin{figure*}[!t]
\centering
\subfloat[Scenario sensitivity.]{
\includegraphics[width=0.48\textwidth]{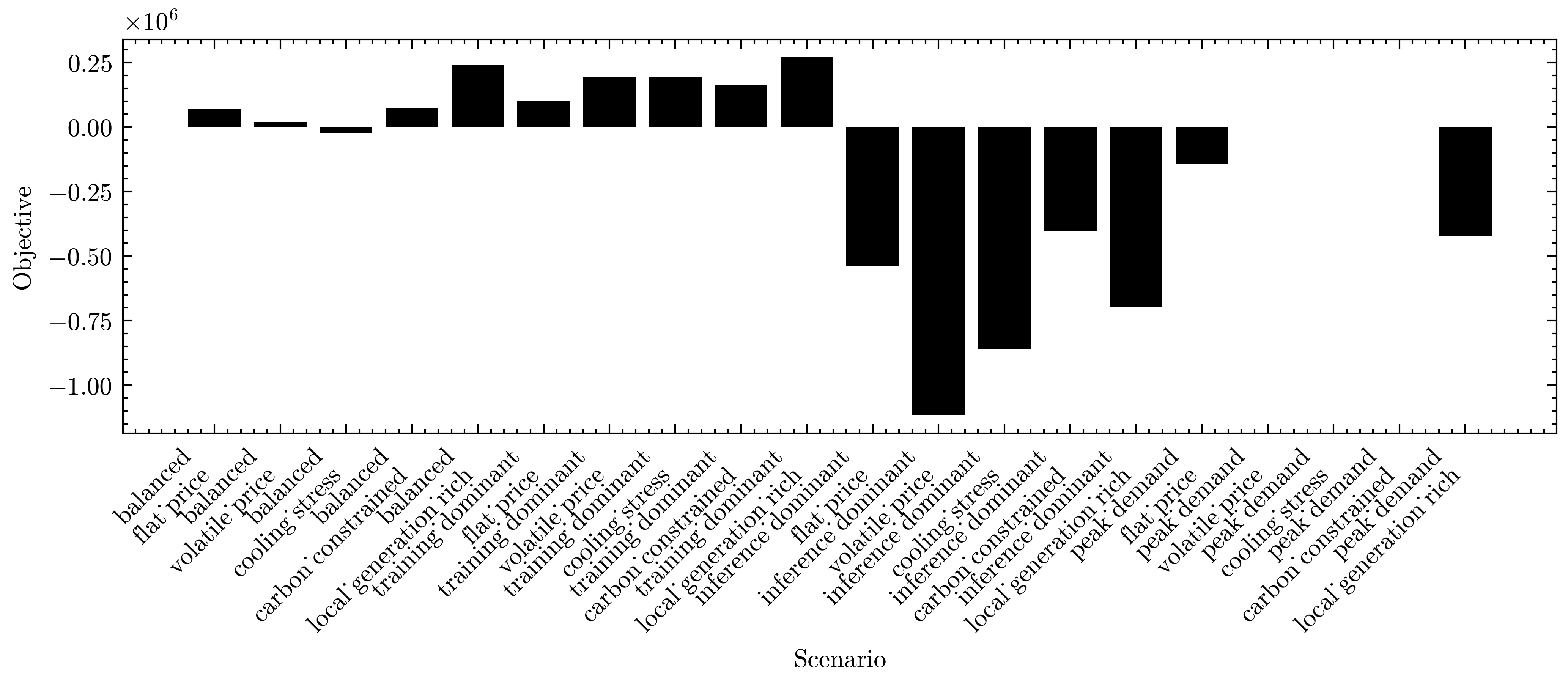}
}
\hfill
\subfloat[Scalability.]{
\includegraphics[width=0.48\textwidth]{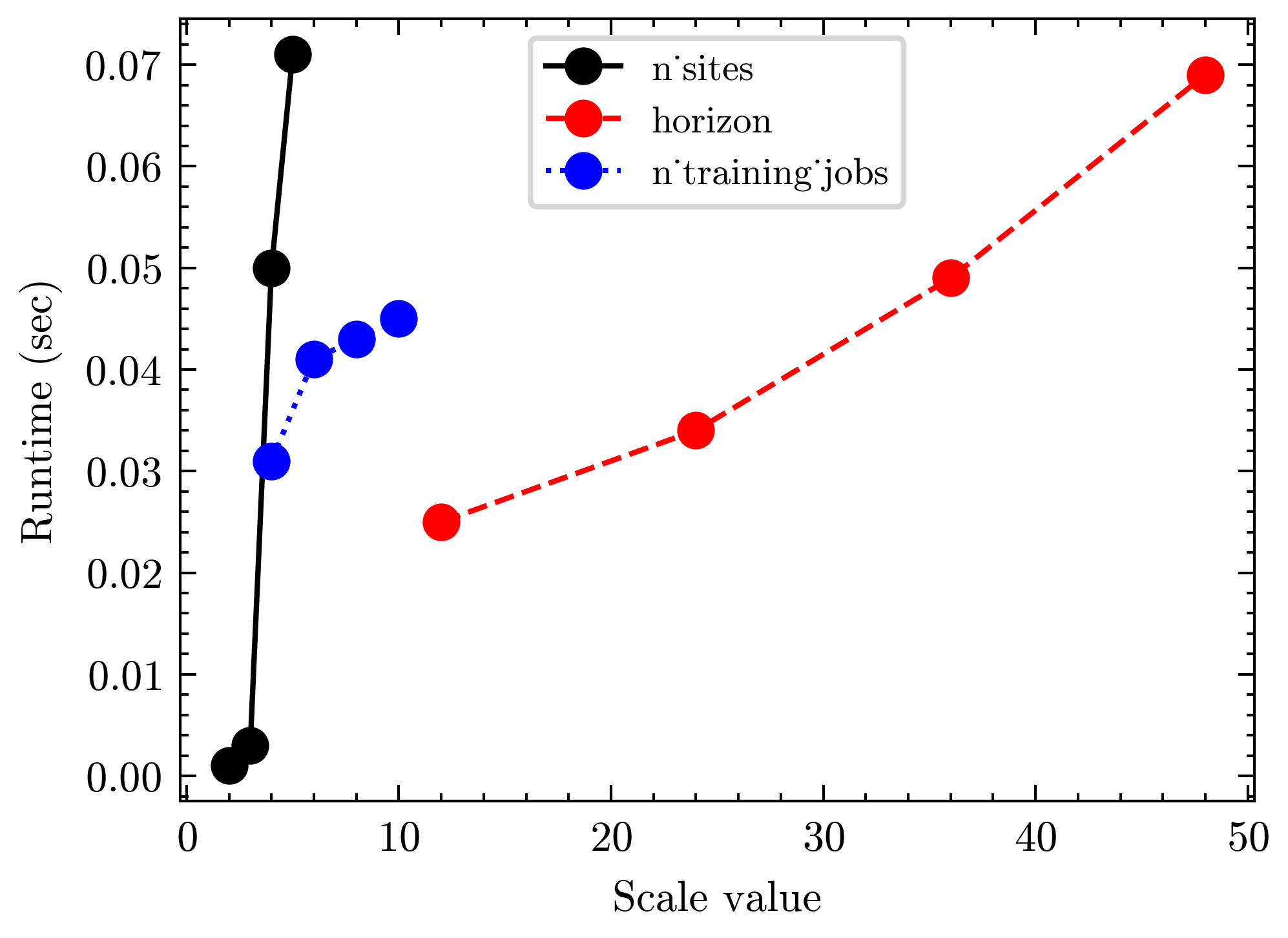}
}
\caption{Sensitivity and scalability of the proposed MILP.}
\label{fig:sens_scale}
\end{figure*}

Fig.~\ref{fig:sens_scale}(a) summarizes the performance of the joint MILP
across workload and energy scenarios. Training-dominant and
local-generation-rich settings are the most favorable, while
inference-dominant and peak-demand settings are more challenging. This
indicates that greater workload elasticity does not automatically imply
easier operation once cooling, procurement, and infrastructure constraints
are jointly considered.

Fig.~\ref{fig:sens_scale}(b) reports runtime as the number of sites, time
periods, and training jobs increases. Feasible medium-scale instances are
solved within fractions of a second under the tested settings. Runtime grows
with horizon length and the number of rigid training jobs, as expected from
the increase in time-indexed binary variables. The site-scaling results also
show that additional spatial diversity can restore feasibility by providing
more workload-placement and energy-balancing options.

Overall, the experiments show that coordinated workload--energy optimization
substantially improves operational performance relative to decoupled
baselines, especially when routing flexibility and storage are available.
They also reveal strong heterogeneity across operating regimes, with
training-dominant and local-generation-rich settings being the most favorable.
\section{Conclusion}
\label{sec:conclusion}

This paper proposed a MILP framework for carbon-aware compute--power
scheduling in geographically distributed AI data centers with microgrid
prosumer capabilities. The model jointly captures rigid training-job
commitment, elastic inference routing, cooling-adjusted power demand, local
generation, battery storage, bidirectional grid interaction, and carbon
constraints.

Experiments on synthetic yet practically motivated instances show that joint
optimization substantially improves operational benefit and reduces emissions
relative to decoupled compute-only and energy-only baselines. The results
also highlight inference-routing flexibility as a major source of value,
battery storage as a useful temporal buffer, and local-generation-rich
settings as particularly favorable operating regimes.

Future work will study uncertainty-aware, online, and decomposition-based
extensions for larger-scale deployments with stochastic demand, renewable
generation, electricity prices, and carbon intensity.






%


\bibliographystyle{IEEEtran}
\bibliography{references}

\end{document}